\documentstyle[12pt]{article}
\begin{document}
\thispagestyle{empty}
\def\cqkern#1#2#3{\copy255 \kern-#1\wd255 \vrule height #2\ht255 depth 
   #3\ht255 \kern#1\wd255}
\def\cqchoice#1#2#3#4{\mathchoice%
   {\setbox255\hbox{$\rm\displaystyle #1$}\cqkern{#2}{#3}{#4}}%
   {\setbox255\hbox{$\rm\textstyle #1$}\cqkern{#2}{#3}{#4}}%
   {\setbox255\hbox{$\rm\scriptstyle #1$}\cqkern{#2}{#3}{#4}}%
   {\setbox255\hbox{$\rm\scriptscriptstyle #1$}\cqkern{#2}{#3}{#4}}}
\def\CC{\mathord{\cqchoice{C}{0.65}{0.95}{-0.1}}}
\def\x{\stackrel{\otimes}{,}}
\def\y{\stackrel{\circ}{\scriptstyle\circ}}
\def\proof{\noindent Proof. \hfill \break}
\def\a{\begin{eqnarray}}
\def\b{\end{eqnarray}}
\def\p{{1\over{2\pi i}}}
\def\Q{{\scriptstyle Q}}
\def\P{{\scriptstyle P}}
\renewcommand{\thefootnote}{\fnsymbol{footnote}}

\newpage
\centerline{\LARGE Lie-Algebraic Characterization}
\centerline{\LARGE of 2D (Super-)Integrable Models}
\vspace{1truecm} \vskip0.5cm

\centerline{\large F. Toppan}
\vskip.5cm
\centerline{Dipartimento di Fisica}
\centerline{Universit\`{a} di Padova}
\centerline{Via Marzolo 8, I-35131 Padova}
\centerline{\em E-Mail: toppan@mvxpd5.pd.infn.it}
\vskip1.5cm
\centerline{\it Talk given in memory of Prof. D.V. Volkov}
\vskip1.5cm
\centerline{\bf Abstract}
\vskip.5cm 
It is pointed out that affine Lie algebras appear to be the natural 
mathematical structure underlying the notion of integrability for 
two-dimensional systems. Their role in the construction and classification
of 2D integrable systems is discussed. The supersymmetric case will be 
particularly enphasized. The fundamental examples will be outlined. 
~\par~\par
\pagestyle{plain}
\renewcommand{\thefootnote}{\arabic{footnote}}
\setcounter{footnote}{0}
\vskip1.5cm
\indent

The integrable hierarchies of differential equations in $1+1$ dimensions
have been widely studied in the last several years both in the physical 
and in the mathematical literature. In the time there has been an 
ever-growing evidence that their relevance should not be confined in the 
realm of pure mathematics, but rather their beautiful mathematical 
structures show themselves naturally when investigating physical
problems.\par
It deserves being mentioned that $2$-dimensional integrable systems 
appear in two different (neverthless related) ways: on one hand we have
the non-relativistic integrable equations in $1$ space and $1$ time 
dimension. The basic example for these systems is the celebrated KdV
equation and the systems of this kind will be referred as of 
KdV-type.\par
On the other hand the second big class of integrable systems is provided
by the relativistic ones in two dimensions, that is the so-called Toda
field theories, whose fundamental example is the even more celebrated
Liouville equation.\par
As discussed later, both such classes are obtainable from one and the 
same  mathematical construction, presenting the affine Lie-algebras
as fundamental ingredient.\par
Before going ahead let us just mention some physical applications of 
both classes of theories, which motivated the interest of physicists
in looking at them. At first the Liouville theory appeared in the 
Polyakov's geometrical attempt in quantizing the bosonic string 
off-criticality \cite{pol}. More recently integrable systems of KdV-type 
were found
associated to physical problems when it was realized they furnished the 
partition functions of the two-dimensional discretized gravity in the 
matrix-models approach (see \cite{dif} for a review). Quite new and rather unexpectedly 
integrable hierarchies appear even associated to $4$-dimensional field 
theories as a sort of an underlying integrable structure of $N=2$  
SuperYang-Mills theories in the Seiberg-Witten framework \cite{sw}.\par
Another way of associating Liouville theory to strings is a rather 
different one. It is based on the 
so-called geometrical approach, greatly developed by the Kharkov's 
group, to 
(classical) strings. In this approach 
the solution for the 
dynamical problem of a string moving on a $2+1$ flat-target is reduced
to the solution of the Liouville equation\cite{omn}. 
It is worth to mention this
point here because it is related to Prof. D.V. Volkov's last work. In 
fact in
\cite{vol} it was constructed the geometrical approach for a 
Green-Schwarz superstring moving in a flat supersymmetric target of
$2+1$ dimensions. Surprisingly, it was found that the equations of 
motions can now be reduced to a supersymmetrized version of the Liouville 
equation, but not the standard one. This result was the basic motivation 
for understanding the situation from purely Lie-algebraic data, work done 
in collaboration with D. Sorokin \cite{st1} and \cite{st2} which will be 
reported later. 
\par
Let us come back now to the mathematical structure of integrable 
systems.
The main difference between KdV-type hierarchies and Toda-type 
hierarchies is due to the fact that the former ones are recovered from a 
single copy of affine algebras, while the latter from two separated 
copies corresponding to the chiral and antichiral sectors respectively.
Let us denote as $J$ the currents valued on a given simple Lie algebra 
${\cal G}$ and generating an affine Lie algebra ${\hat{\cal G}}$.\par 
WZNZ 
models have dynamics expressed by group-manifold elements $G$, while the 
currents are defined as $J= {\partial }G\cdot G^{-1}$ and
${\overline J} = G^{-1}{\overline \partial G}$ (the antichiral one)
respectively and satisfy free-equations. Neverthless in both cases, 
non-trivial equations are obtained by imposing constraints on the affine 
currents $J$ (and ${\overline J}$ in the second case). Such constraints
arise in two different ways, either as hamiltonian constraint, or as 
coset constraints. The first case correspond to the Dirac's theory of 
constraints, while the latter simply means that the dynamical quantities 
should have vanishing Poisson brackets with respect to some Kac-Moody 
subalgebra.\par 
In the basic example of $ sl(2)$ (or $A_1$ algebra) these 
two constructions (denoted as ${ a)}$ and ${ b)}$) lead respectively 
to (for non-relativistic, type-$1$ systems, and relativistic, type-$2$ 
systems) to:
\\
${ 1 a)}$ KdV equation;\\
${ 1 b)}$ NLS equation;\\
${ 2 a) }$ Liouville equation;\\
${ 2 b)}$ Witten's $2D$'s black hole.\par
The arising of integrable hierarchies from constrained affine Lie 
algebras is particularly important because it provides the tools towards 
a classification of all hierarchies.\par
Let us now discuss the case of supersymmetric extensions of bosonic 
hierarchies. The interest in such extensions should not be thought being 
limited to the super-physics program (and more specifically to 
superstrings), instead a wider range of applications arises. For 
instance it is well-known that new bosonic hierarchies can arise when 
only the bosonic (or more generally the non-supersymmetric sector) of 
theories based on supergroups and superalgebras is considered.\\
One of the main sources of interest in investigating supersymmetries 
morever coincide with large $N$-supersymmetric extensions, because it 
correspond to a sort of ``unification" or ``grandunification" of known 
hierarchies. It happens indeed that seemingly unrelated bosonic or lower 
supersymmetric ($N=1,2$) hierarchies turn out to be a different 
manifestation of a single ``unifying" large-$N$ supersymmetric hierarchy. 
\par
Until recently it was commonly believed that supersymmetric hierarchies 
could be produced only from affinization of a particular kind of 
superalgebras. It must be explained that superalgebras, just as ordinary 
algebras, can be expressed through their Dynkin's diagrams which refer 
to their simple roots. However, since superalgebras admit two kinds of 
generators, even and odd, the simple roots could be either fermionic or 
bosonic. It was thought that only the special class of superalgebras admitting 
purely fermionic simple roots could provide supersymmetric integrable 
models. The reason for that was based on an argument related to the 
Drinfeld-Sokolov approach to integrable systems (based on simple roots).
Since the basic derivative operator for supersymmetric theory is 
fermionic it was thought that the only consistent way to construct a 
fermionic matrix-Lax pair implied the use of the fermonic simple roots.
It appeared at first as a surprise in \cite{bd} that the supersymmetric
version of the Non-Linear-Schr\"odinger equation admits a Lax pair based 
on $sl(2)$ instead of $osp(1|2)$ as expected. In \cite{top} it was shown
that such supersymmetric equation admits a natural interpretation in
terms of a coset construction based on the supersymmetric affinization 
of ${\hat{sl(2)}}$ (the tower of hamiltonian densities has vanishing 
Poisson brackets w.r.t. the supersymmetric ${\hat u(1)}$ subalgebra).\par
At this stage it was clear that, at least for coset construction, it was 
perfectly acceptable to recover supersymmetric integrable models from 
any bosonic or super Lie algebra. \par
More recently, in collaboration with D. Sorokin \cite{st1} and 
\cite{st2}, we have shown how it is possible to bypass the requirement 
of purely fermionic Dynkin's diagram-type supealgebras even in the 
case of hamiltonian constraint (the case $a)$ in the previous 
classification). For lack of space I cannot describe here our method and 
I refer to the cited papers for details. Let me just point out that
the resulting hierarchies (in the Toda-type construction) are 
superconformally invariant, with the supersymmetry realized 
non-linearly and spontaneously broken. Such kind of systems give us 
automatically a
non-standard Sugawara realization of the superconformal stress-energy 
tensor which involves fermionic $b-c$ systems of weight 
$(-{\textstyle {1\over 2}},{\textstyle{3\over 2}})$.
The simplest example of this kind is obtained in terms of the $sl(2)$
algebra. It corresponds to the non-standard superLiouville equation
found in \cite{vol} expressing the dynamics of a Green-Schwarz 
superstring on a $2+1$ Minkowski flat target.\par
In the last part of my talk I wish to introduce some new results,
found in collaboration with E. Ivanov and S. Krivonos, concerning the 
large-$N$ supersymmetric extension of integrable hierarchies (and their 
relation to affine algebras) \cite{ikt}. A bosonic algebra such as 
$sl(2)\oplus u(1)$ 
turns out to be the basic structure underlying the $N=4$ KdV 
hierarchy.\par  
Indeed the following features hold. Algebras admitting quaternionic 
structure have been classified \cite{cla}. They are thought to be 
related to $N=4$ theories. Indeed the simplest non-trivial case 
($sl(2)\oplus
u(1)$, the abelian $u(1)^{\oplus 4}$ should be ruled out for our 
purposes) is such that their supersymmetric affinization admits a
global $N=4$ structure (realized by non-linear transformations). 
Moreover an infinite number of $N=4$ hamiltonians in involution 
associated to the above superaffine algebra as Poisson 
bracket structure, can be found. The resulting integrable hierarchy
can be denoted as $N=4$ NLS-mKdV hierarchy because different constraints
compatible with the equations of motion lead respectively to the 
$N=2$ mKdV and to the $N=2$ NLS equation.\par
The $N=4$ hierarchy is not only supersymmetric but even 
$N=4$ superconformal because the Sugawara construction applied to the 
superaffine ${\widehat{sl(2)\oplus u(1)}}$ leads to a closed algebraic
structure, where the $N=4$ transformations are linearized, which 
corresponds to the so-called minimal $N=4$ version of the SuperConformal 
Algebra (expressed in terms of three bosonic spin $1$ $N=2$ superfields,
one bosonic, one chiral and one antichiral). With respect to the
generators of the $N=4$ SCA the equations of motions are closed and 
coincide with the equations of the $N=4$ KdV hierarchy \cite{kdv}.
\par
Therefore even large-$N$ supersymmetric hierarchies find an 
interpretation in terms of (super-)affine Lie algebras. This result is 
particularly important because it paves the way towards an 
understanding and a Lie-algebraic classification of all $N=4$ 
hierarchies.

~\par~\par
{\bf Acknowledgements:} \par
~\par
I wish to express my gratitude to the organizers of the conference 
in memory of Prof. D.V. Volkov for their kind invitation to give a talk.\par
~\par
~\par

\end{document}